# Electronic Structure of Manganese Phthalocyanine Modified via Potassium Intercalation: a Comprehensive Experimental Study

Francisc Haidu[1], Ovidiu D. Gordan[1], Dietrich R. T. Zahn[1], Lars Smykalla[2], Michael Hietschold[2], Boris V. Senkovskiy[3,4], Benjamin Mahns[5], and Martin Knupfer[5]

[1] Semiconductor Physics, Technische Universität Chemnitz, D-09107 Chemnitz, Germany
[2] Solid Surfaces Analysis Group, Technische Universität Chemnitz, D-09107 Chemnitz
[3] Germany, Institute of Solid State Physics, Dresden University of Technology, D-01062 Dresden, Germany
[4] St. Petersburg State University, 198504 St. Petersburg, Russia
[5] Electronic and Optical Properties Department, IFW Dresden, D-01171 Germany

## Abstract

Potassium (K) intercalated manganese phthalocyanine (MnPc) reveals vast changes of its electronic states close to the Fermi level. However, theoretical studies are controversial regarding the electronic configuration. Here, MnPc doped with K was studied by ultraviolet, X-ray, and inverse photoemission, as well as near edge X-ray absorption fine structure spectroscopy. Upon K intercalation the Fermi level shifts toward the lowest unoccupied molecular orbital filling it up with donated electrons with the appearance of an additional feature in the energy region of the occupied states. The electronic bands are pinned 0.5 eV above and 0.4 eV below the Fermi level. The branching ratio of the Mn $L_3$ and $L_2$ edges indicate an increase of the spin state. Moreover, the evolution of the Mn $L$ and N $K$ edges reveals strong hybridization between Mn 3d and N 2p states of MnPc and sheds light on the electron occupation in the ground and n-doped configurations.

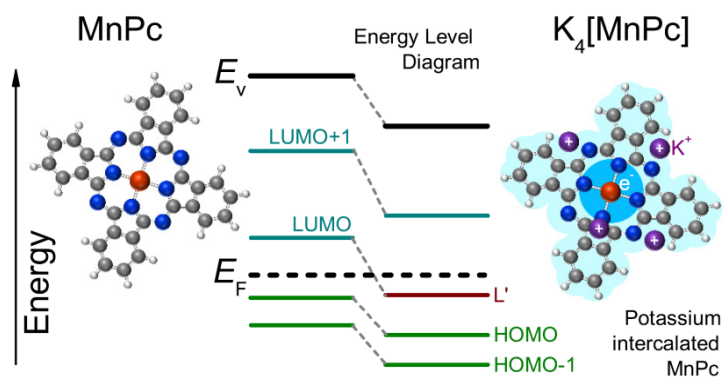

# Introduction

The cumulative knowledge on the electronic properties of organic semiconductors is a key element in optimizing and developing new molecular devices, such as organic field effect transistors,[1-4] organic light-emitting devices[5-7] and thin film organic photovoltaic cells.[8,9] Moreover, the energy band alignment at metal–organic semiconductor interfaces[10] plays a key role in the emerging field of organic spintronics, which combines the potentials of spintronics[11] and organic electronics.[12] Besides low production costs, high flexibility, and never ending opportunities in molecular design, organic molecules have long spin-coherence lifetimes which makes them promising candidates for future spintronics applications.[13-17]

The aim of integrating the spintronic functionality of single molecules or thin molecular films into nanostructured devices requires the control over the molecular spin ground state and its spin filtering properties. The phthalocyanine (Pc) class of molecules has very good prospects toward spintronic applications.[14,15,17] Besides the electronic and optical properties, the spin state of the Pc molecule is highly tunable by adjusting the peripheral ligands or by exchanging the central metal ion.[14,17-19]

Among the transition metal phthalocyanine (TMPc) series, MnPc is one of the most interesting due to its specific magnetic properties, in the bulk,[20,21] as well as in thin films,[22] at low temperatures. This molecule is considered a very good candidate for spintronic devices due to its intermediate-spin $S = 3/2$ ground state [20,23] and nearly perfect spin filtering properties.[24] The magnetism of a single MnPc molecule was addressed by scanning tunneling microscopy demonstrating a technologically feasible way to manipulate a single spin.[25]

The spin state of the MnPc molecule can be decreased to $S = 1$ by adsorption on a Bi(110) surface.[26] It can even be decreased to the low-spin configuration of $S = 1/2$ by reaction with CO,[26] or $O_2$,[27] or by ligation with pyridine.[28] Another recent discovery was of the MnPc$^{\delta+}$ / $F_{16}$CoPc$^{\delta-}$ charge transfer dimer which has a total spin state of $S = 2$.[29,30]

A very effective and elegant method for tuning the electronic, optical, and spin properties of the MnPc molecule is by alkali metal (AM) intercalation. In Refs. 17,23 it was reported that the high-spin state of $S = 5/2$ was obtained upon intensive AM doping. This is of great significance for spintronic applications. Moreover, also the electrical conductivity of the MnPc organic molecular film was highly increased with this method.[31,32] Through the K intercalation process of MnPc three stable doped phases were obtained, unlike for other Pcs which only showed two electronically stable phases.[33]

Photoemission spectroscopy (PES) and X-ray absorption spectroscopy (XAS) studies on AM intercalated Pc molecules are fairly common: such as $H_2$Pc,[34] ZnPc,[35,36] CuPc,[17,37-41] NiPc,[17] CoPc,[42] and FePc.[17,43,44] However, AM intercalated MnPc was only recently studied by XAS and density functional theory calculations.[17]

In this study MnPc was investigated in the pristine and K intercalated states by combined ultraviolet photoemission spectroscopy (UPS) and inverse photoemission spectroscopy (IPS), by X-ray photoemission spectroscopy (XPS), as well as by near edge X-ray absorption fine structure (NEXAFS) spectroscopy.

While inorganic semiconductors have the transport band gap energetically very close to the optical band gap and it can thus be determined by optical absorption spectroscopy or spectroscopic ellipsometry measurements, organic semiconductors present a large mismatch between the optical and transport band gaps which is caused by the strong excitonic effects in organic molecules.[45,46] As a consequence, the optical band gap can be up to 1.4 eV smaller than the transport bandgap.[45] Their difference represents the exciton binding energy. Therefore, for the correct transport band gap determination, the combination of direct and inverse photoemission measurements was employed. Moreover, they provide information on the evolution of energy levels upon K intercalation.

With XPS measurements the relative position of the $K^+$ ions with regard to the MnPc molecular body was determined. Furthermore, from the NEXAFS spectra the positioning of the donated electrons was identified.

The ground state electronic configuration of MnPc is still under debate in literature (see e.g. Refs. 17, 47, 48). This work aims to aid theoretical calculations by providing a comprehensive experimental study of MnPc in its pristine and n-doped configurations.

## Experimental

### Sample Preparation and Measurement

MnPc molecules (Strem Chemicals, Inc., 97% specified purity) were sublimed from Knudsen cells by organic molecular beam deposition with a constant rate of 0.2 nm/min at the temperature of (380 ± 10) °C. The amount of sublimed material was monitored by the shift in the oscillation frequency of a calibrated quartz crystal microbalance. The base pressure in the preparation chambers was lower than $5\times10^{-10}$ mbar and did not exceed $1\times10^{-8}$ mbar during the sublimation. Pure Co foils (Alfa Aesar, 99.997% Puratronic) were employed as substrates. The surface impurities were removed by $Ar^+$ ion bombardment.

The samples were composed of 10 nm MnPc deposited onto Co. This was thick enough that the substrate influence was minimized[10] and thin enough that no charging effects were present.[49] Subsequently, potassium was evaporated at a constant rate (heating current of 7.0 A) from a dispenser (SAES Getters S.p.A, Italy) mounted 10 mm in front of the sample. The amount of evaporated K was monitored and calibrated by XPS.

All the measurements were performed in ultra-high vacuum (UHV) conditions with the base pressures in the low $10^{-10}$ mbar range.

The UPS and IPS systems are attached to the same UHV chamber. As an excitation source for the UPS experiments a He gas discharge lamp was employed which was operated with the He I (21.2 eV) excitation line. The photo-emitted electrons were analyzed with a hemispherical analyzer. The IPS system operates in the isochromatic mode with a low energy electron gun as the excitation source and a Geiger-Müller (GM) tube as the photon detector. The current density ($< 10^{-6}$ A/cm$^2$) was low enough not to damage the organic molecules. The GM counter operates as a fixed energy photon detector (10.9 eV) provided by the Ar (10 mbar) and ethanol (2 mbar) gas mixture and the MgF$_2$ entrance window. The energy resolution of the UPS and IPS setups was determined as 0.2 eV and 0.5 eV, respectively. Clean Au and Co foils were used for the determination.

XPS and NEXAFS measurements were performed at the Russian-German beamline at the BESSY II synchrotron radiation facility, Helmholtz-Zentrum Berlin. The XPS experimental station is equipped with a SPECS PHOIBOS 150 analyzer with a 2D CCD detector. In both XPS and UPS measurements the photo-emitted electrons along the normal to the sample surface were analyzed. The XPS acquisition parameters, i.e. excitation energy, cross section, energy resolution, and pass energy, for each core level, i.e. Mn 2p, N 1s, K 2p, C 1s, and the valence band (VB), are given in Table 1. For the NEXAFS experiments the total electron yield (TEY) signal was recorded. The photon flux was determined from the photo-current on a clean gold grid placed in the beam path in front of the sample. The signal was employed in the TEY spectrum normalization.

**Table 1:** XPS acquisition parameters.

| Energy level | Excitation energy (eV) | Cross section (Mb)[a] | Energy resolution (eV) | Pass energy (eV) |
|---|---|---|---|---|
| Mn 2p | 750 | 1.2[b] | 0.30[c] | 30 |
| N 1s | 500 | 0.45[b] | 0.23[c] | 20 |
| K 2p | 500 | 1.0[b] | 0.23[c] | 20 |
| C 1s | 500 | 0.3[b] | 0.23[c] | 20 |
| VB | 150 | – | 0.15[c] | 20 |

[a] Mb (mega barn) represents a unit of area, 1 Mb = $10^{-22}$ m$^2$; [b] Ref. 50; [c] Ref. 51.

## Data Analysis

The valence band spectra are plotted in Figs. 1(a), (b), and (c) with decreasing binding energy (BE). Following the same convention, the inverse photoemission spectra [Fig. 1(d)] continues on the same energy scale with negative values as all the spectra are plotted with respect to the Fermi level ($E_F$ = 0 eV).

The energy position of the vacuum level ($E_v$) is determined by subtracting the excitation energy from the secondary electron cut-off positions. For the Co substrate the absolute value of this quantity is its work function ($\Phi = -E_v$). The features closest to the Fermi level in the UPS–IPS measurements on the pristine MnPc film are the highest occupied molecular orbital (HOMO) and the lowest unoccupied molecular orbital (LUMO), respectively.

The UPS, IPS, and XPS data were fitted using the Unifit2010 data analysis software.[52] The features in the HOMO and LUMO regions (UPS–IPS) were fitted with Gaussian line shapes. A third order polynomial function was implemented as background in the analysis procedure.[10,49,53] The onset of the HOMO features were determined as the intercept of the tangent to the fitted Gaussian with the x-axis. Due to the poor energy resolution of the IPS technique the experimental broadening ($Br_{ex}$ = 0.5 eV) of the LUMO features has to be taken into account. Thus, a deconvolution of the measured peaks has to be performed according to the formula:

$$Br_L^2 = Br_{me}^2 - Br_{ex}^2, \qquad (1)$$

where $Br_{me}$ represents the full width at half maximum (FWHM) of the measured peak and $Br_L$ is the FWHM of the peak employed to determine the LUMO onset position.

The core level spectra were analyzed using Voigt line shapes for the peaks and Shirley[54] type backgrounds. The limited lifetime of the core-hole state produces a Lorentz type broadening while the measurement setup induces a Gaussian broadening. The convolution of the two profiles gives the Voigt lineshape.[55] To compensate the error induced by the monochromator in the energy scale, the XPS data was corrected by identifying the Au 4f peak positions on a clean Au foil.

The number of K atoms per MnPc molecule ($N_K$) was determined by XPS measurements according to the formula:

$$N_K = N_C \frac{A_K}{A_C} \frac{\sigma_C}{\sigma_K}, \qquad (2)$$

where $N_C$ = 32 is the number of C atoms per molecule, $A_K$ and $A_C$ represent the K 2p and C 1s peak areas, respectively, while $\sigma_K$ and $\sigma_C$ are their respective cross sections (see e.g. Table 1). Note, that the transmission function of the analyzer was neglected in Eq. (2) as its value was constant in the short energy range (~10 eV) where the two core levels were measured. By plotting $N_K$ against the K evaporation time the calibration curve was obtained. The slope of the linear fit yields a doping rate $r$ = (0.15 ± 0.02) min$^{-1}$ as K atoms per molecule per minute. This value was employed to determine the amount of evaporated K in the UPS-IPS experiments.

NEXAFS spectra were recorded at the Mn $L_{2,3}$, N $K$, and C $K$ edges of the pristine and K intercalated MnPc films. All Mn $L_{2,3}$ spectra were normalized to 0 at the pre-edge region and to 1 at the region between the $L_3$ and $L_2$ edges. Both N $K$ and C $K$ edges are composed of two regions: at lower energies the features belong to transitions from the

1s core level to π* antibonding orbitals while at higher energies they belong to transitions from the 1s core level to σ* antibonding orbitals. All N *K* and C *K* edges spectra were normalized to 0 at the pre-edge region and to 1 at the spectral tale (above the 1s → σ* features). More information on NEXAFS spectroscopy can be found in Ref. 56.

## Results and Discussions

### Ultraviolet Photoemission Spectroscopy and Inverse Photoemission Spectroscopy

The impact of K intercalation on the occupied and unoccupied electronic states of MnPc was studied by the combined UPS and IPS study. On a clean Co foil 10 nm MnPc was sublimed and consecutively intercalated with K. Figure 1 presents UPS and IPS measurements on the clean substrate, the MnPc film, and the K intercalated layer as a function of K dosage. The amount of K atoms per MnPc molecule was calibrated by XPS and asserted as a function of K evaporation time. The error in $N_K$ determination is 15% and is not specified in the legend.

The HOMO and LUMO regions of the spectra [Figs. 1(c) and (d)] were fitted according to the description presented above and the resulting components are plotted in the figure. The features in the LUMO region were deconvoluted by employing Eq. (1) and outlined with gray dashed lines in Fig. 1(d). Furthermore, the onset positions of the HOMO (H), HOMO−1 (H−1), LUMO (L), and LUMO+1 (L+1) states were marked in Figs. 1(c) and (d). With increasing amount of K the LUMO state is filling up with electrons and appears as an additional occupied state ascribed by L' in the valence band region. At the same time the intensity of the LUMO feature decreases until it fully disappears for high K dosage.

It should be mentioned that the features in the UPS-IPS data were smeared out with increasing amount of K in the molecular film. This broadens the features, especially in the IPS part, as was observed also in other works (see e.g. Refs. 35,57).

The onset energy positions of the above mentioned quantities are given in Fig. 2(a) with regard to the Fermi level ($E_F$). Up to the K dosage of $N_K$ = 0.9 the HOMO and HOMO+1 have a slight shift toward $E_F$ (i.e. lower BE) while the other features shift toward higher BE. Such small shifts were observed at low K dosage. Orbitals with antibonding character fill up with the donated 4s electrons of the K atoms modifying the whole electronic configuration and the molecular bond lengths. Moreover, bond lengths can also be influenced by the $K^+$ ions which reside close to the MnPc molecules. Theoretical calculations determined different electronic ground state configurations, with different HOMO−LUMO gap, as a function of bond lengths in the MnPc molecule.[48]

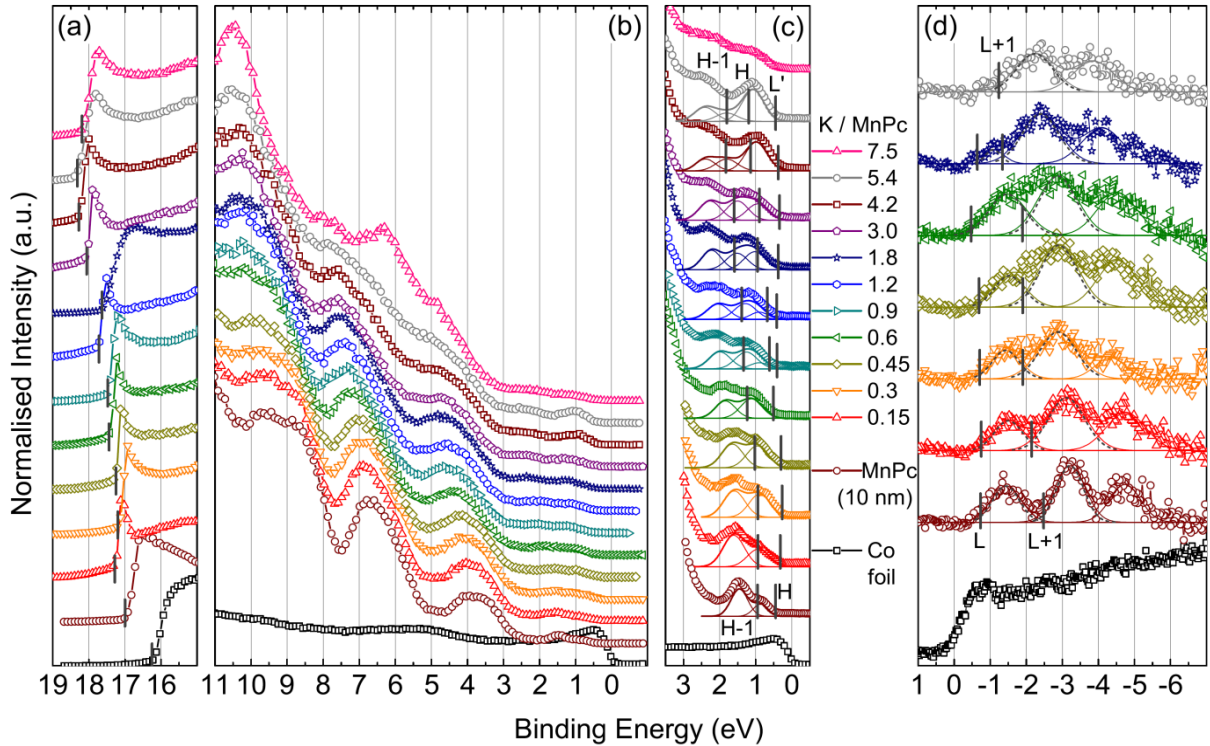

**Figure 1:** Evolution of the UP (a), (b), and (c) and IP (d) spectra of K intercalated MnPc film deposited on Co foil. The UP spectra are divided in three regions: secondary electron cut-off (a), overview of the valence band (b), and the HOMO region (c). For clarity, the IP spectra are background subtracted, except for the Co foil spectrum (d). The secondary electrons cut-off positions, the onset positions of the HOMO (H), HOMO−1 (H−1), filled LUMO state (L'), LUMO (L), and LUMO+1 (L+1) are marked with vertical bars.

In conclusion, occupied and unoccupied molecular orbitals present slightly different shifts by increasing the K amount.[58]

Above a K dosage of $N_K$ = 0.9 all bands shift toward higher BE. For the fully intercalated film ($N_K$ > 4) the shifts reach as much as 1 eV. The exceptions are the LUMO and L' which do not cross $E_F$.[58] Due to this pinning of the two bands, or Fermi level pinning,[35,39,41,57,59–66] the molecular thin films do not present metallic properties from the electronic point of view and remain semiconducting.[67] However, it was previously shown by electrical measurements that the conductivity increases dramatically by AM intercalation.[31] The reason could be the reduction of the transport bandgap upon doping from (1.2 ± 0.3) eV to (0.9 ± 0.3) eV (see e.g. Fig. 2). Moreover, for the fully doped MnPc ($N_K$ > 4) the LUMO state disappears leading to the decrease of the current as observed in Ref. 31. Certainly without the LUMO state the bandgap increases again.

The energy level diagram of the pristine and fully doped ($N_K$ = 4) MnPc film is plotted in Fig. 2(b). While other studies report a rigid shift of both HOMO and LUMO bands upon doping (see e.g. Ref. 57) there are small differences to be observed between the shifts of occupied and unoccupied electronic states. While H and H−1 positions change by ∼0.8 eV the L+1 state, on the other hand, has a total shift of ∼1.2 eV. Moreover, in the following it will be shown that the core levels have an even lower energy shift with the amount of intercalated K.

As mentioned above, the remaining LUMO and the L' states are pinned at 0.5 eV above and 0.4 eV below the Fermi level [Fig. 2(b)]. The pinning energy positions depend on many factors: the substrate work function,[68–70] the thickness of the organic film,[71] the doping type (n- or p-type),[35,59] or the doping material.[35,59,72] The reason for the $E_F$ pinning are gap states which are commonly attributed to structural defects.[73,74]

## X-ray Photoemission Spectroscopy

In Fig. 3 the core level spectra of the pristine and K intercalated MnPc films are presented with increasing amount of K from top to bottom. At high binding energy the Mn $2p_{3/2}$ and Mn $2p_{1/2}$ features are plotted in Fig. 3(a). The doublet peaks were coupled during the fitting procedure, however, for clarity only the $2p_{3/2}$ peaks are presented in detail. In the MnPc molecule the manganese central ion has the oxidation state of $Mn^{2+}$ (Ref. 75). This is plotted with the red hatched features in the Mn $2p_{3/2}$ fitted spectra and assigned as Mn(II). In the pristine form the Mn(II) peak has a shoulder at higher BE plotted with blue hatched area and designated as *multi* in the figure. This extra feature is a consequence of the multiplet effects.[17,47] The multiplet splitting has an even larger effect on the Mn $L_{2,3}$ edge NEXAFS spectral shape.

The main electronic configuration of MnPc in the ground state is still not well understood and the way the 3d levels are populated remains under debate.[48] Kroll et al. pointed out the equal possibility of two configurations: $b_{1g}^0 a_{1g}^1 e_g^3 b_{2g}^1$ and $b_{1g}^0 a_{1g}^1 e_g^2 b_{2g}^2$, where $b_{1g}$, $a_{1g}$, $e_g$, and $b_{2g}$ represent the 3d shell orbitals with increasing BE in the given order.[47] The superscripts in the configurations represent the number of electrons per orbital. Recently Stepanow et al.[17] pointed out the sole configuration of $b_{1g}^0 a_{1g}^1 e_g^2 b_{2g}^2$ while Brumboiu et al.[48] determined the $b_{1g}^0 a_{1g}^1 e_g^3 b_{2g}^1$ configuration as the ground state electron configuration.

As the amount of K in the MnPc film increases the *multi* peak decreases in intensity and has an insignificant contribution for $N_K > 1.8$ [Fig. 3(a)]. This means that the ground state electron configuration of the pristine state changes and accommodates donated electrons upon K intercalation. The third feature (*sat*) at 6 eV higher BE, plotted with the green hatched area in Fig. 3(a), was assigned to a further satellite. It is better visible for the fully doped sample ($N_K$ = 5.7). The assignments of the Mn $2p_{3/2}$ features were made according to Refs. 47, 76, 77.

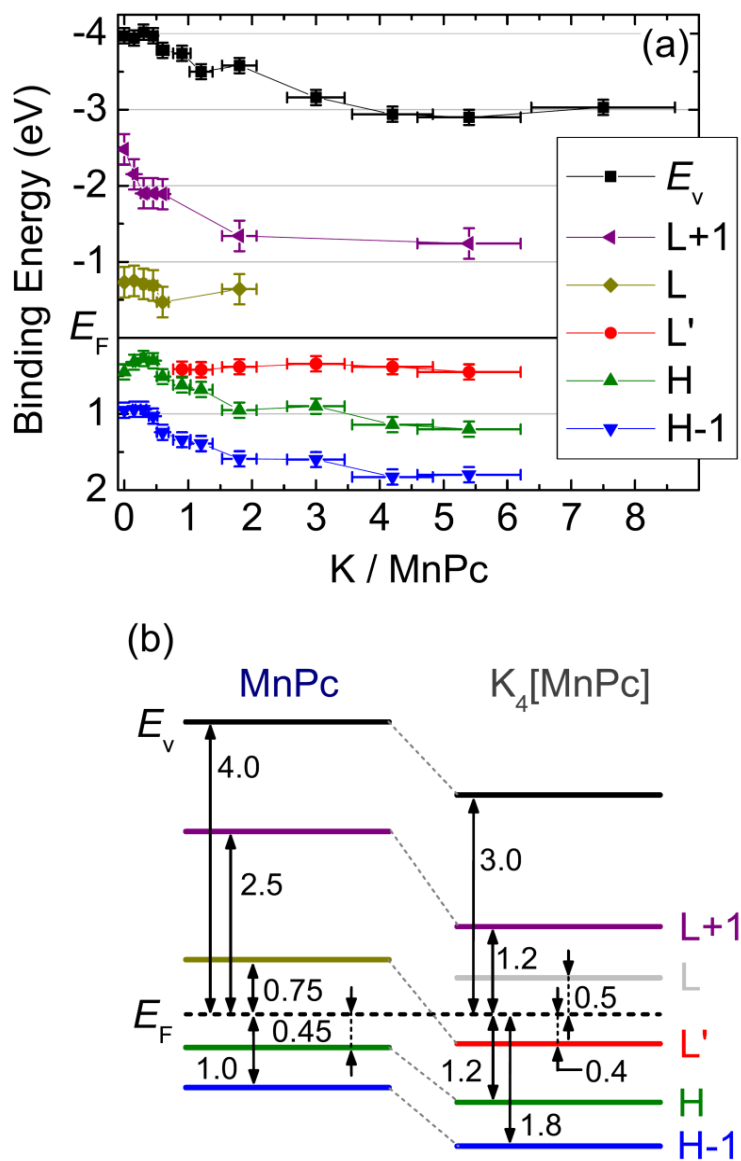

**Figure 2:** (a) The onset positions of the frontier molecular orbitals (quantities determined in Fig. 1) as a function of K dosage. The vacuum level ($E_v$) positions are determined from the secondary electron cut-off positions. (b) Energy level diagram of the occupied and unoccupied electronic states of MnPc before and after K intercalation.

Further, the N 1s data is presented in Fig. 3(b) for the pristine and K intercalated MnPc film. The pristine material presents two main features $N_M$ and $N_P$ plotted with red and blue hatched areas, respectively. The two peaks are the result of two nitrogen species which are schematically represented in Fig. 3(e): $N_M$ is connected to the central Mn atom while $N_P$ bridges the pyrrole rings.43 Their shake-up satellites were assigned as $SN_M$ and $SN_P$, respectively.

Influenced by the K+ ions the $N_P$ feature decreases in intensity [Fig. 3(b)]. An extra feature, assigned as $N_R$ (green hatched area), rises at lower BE. Its shake-up satellite was labeled as SNR. As the intensity of the NR component becomes even larger than NM for the fully doped MnPc film ($N_K = 5.7$), it can be deduced that also NM atoms are affected by the $K^+$ ions. Thus, although K atoms preferentially occupy positions between the benzene rings (close to $N_P$),[43] above a certain K concentration ($N_K > 1.8$) positions closer to the metal center start to be occupied. The sites preferentially occupied by the K⁺ ions are schematically sketched in Figs. 3(f), (g), and (h) for the K dosage of $N_K = 1$, $N_K = 2$, and $N_K = 4$, respectively.

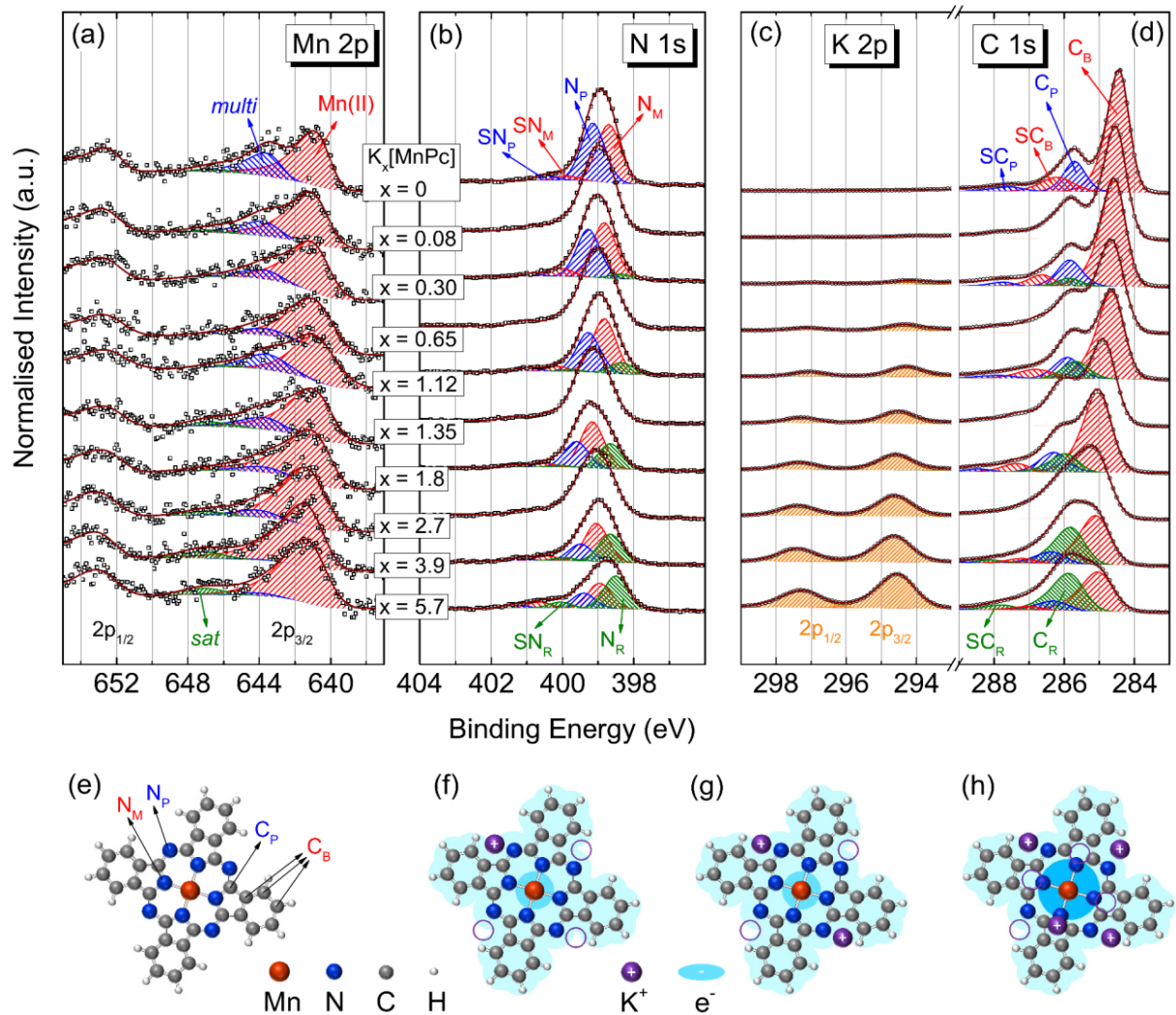

**Figure 3:** MnPc core level spectra as a function of K dosage. The component analysis spectra of the Mn $2p_{3/2}$ (a), N 1s (b), K $2p_{3/2}$ and K $2p_{1/2}$ (c), and C 1s (d) regions. The amount of K increases from top to bottom as indicated in the legend between panels (a) and (b) and holds for each core level spectrum. For clarity some of N 1s and C 1s core level spectra are plotted without the fitted component features. (e) Schematic representation of the pristine MnPc molecular structure with the constituent atoms and the labeling of different N and C species. Schematic representation of the K intercalated molecule with the preferred positioning of the $K^+$ ions and of their donated electrons ($e^-$) as a function of the K amount: $N_K = 1$ (f), $N_K = 2$ (g), and $N_K = 4$ (h). The probability of the donated electrons (or $e^-$ cloud) position is represented by shades of blue: darker area means more $e^-$. The purple circles in (g) and (h) illustrate other possible locations for the $K^+$ ions.

In Fig. 3(c) the evolution of the K 2p spectra is further analyzed. These were fitted with a doublet representing the $2p_{3/2}$ and $2p_{1/2}$ components with the peak area ratio of 2:1, respectively. Finally, the components of C 1s are presented in Fig. 3(d). Pristine MnPc shows two main features $C_B$ and $C_P$ plotted by hatched red and blue areas, respectively. They originate from carbon atoms of the benzene ($C_B$) and of the pyrrole rings ($C_P$) which are schematically shown in Fig. 3(e). The $C_B$ to $C_P$ peak area ratio is 3:1. Their shake-up satellites are ascribed by $SC_B$ and $SC_P$. With the increase of the K dosage an additional peak arises between the $C_B$ and $C_P$ components. It is designated as $C_R$ (with its shake-up satellite $SC_R$) and depicted in Fig. 3(d) with a green hatched area. $C_R$ results from both benzene and pyrrole C atoms influenced by the K⁺ ions.

The shake-up satellites of N 1s and C 1s core levels for pristine MnPc are at ∼ 1.25 eV and ∼ 1.8 eV higher BEs than the main peaks. The C 1s related satellite main peak splitting of MnPc has a value close to the one of CuPc (between 1.7 eV and 2.0 eV),[10] however the N 1s related satellite main peak splitting of MnPc is much smaller than for CuPc (1.8 eV).[10] Moreover, the relative position of the N 1s related shake-up satellites with regard to the main peaks for MnPc (∼ 1.25 eV) can be directly related with the transport bandgap of MnPc, $E_t = (1.2 \pm 0.3)$ eV determined by the combined UPS–IPS measurements.[49] This is in very good agreement with the fact that the HOMO is composed of hybridized Mn 3d orbitals with the surrounding N ligand orbitals.[78] On the other hand, the C 1s shake-up satellites are related rather to the HOMO−1 (π nature ligand) electronic states. Therefore, they present a larger splitting with regard to the main features (∼ 1.8 eV).[58] Another effect of donated electrons is the shift of the Fermi level toward the vacuum level and, thus, also of all the core levels toward higher energy positions. In Fig. 4 the relative shifts of the main core levels are plotted against the amount of K in the MnPc layer. The energy scale is flipped over for direct comparison with the energy level diagram from Fig. 2(b). The magnitude of the shifts is related to the arrow in the plot (Fig. 4). Its length represents 0.5 eV and its direction shows the increase in BE. An average of all the core level shifts is given by the red dashed line. A small shift for $N_K < 1$ can be deduced, while for $N_K > 1$ a larger one is observed with saturation above $N_K = 4$. The total shift is less than 0.5 eV, thus, half the value determined for the valence band region (∼ 1.0 eV).

Additional information on the 3d electronic states can be gathered by performing valence band PES measurements at higher excitation energy.[79] In Fig. 5 the evolution of the VB with increasing amount of K is presented in solid lines. The spectra were recorded at an excitation energy of 150 eV. The features denoted by H−1, H, and L' represent the HOMO−1, HOMO, and filled LUMO states, respectively. They present similar behavior as in the UPS–IPS experiments [Fig. 1(c)]. For comparison some of the UPS spectra are presented in Fig. 5 with rectangular symbols. Only UPS spectra with similar $N_K$ values are given.

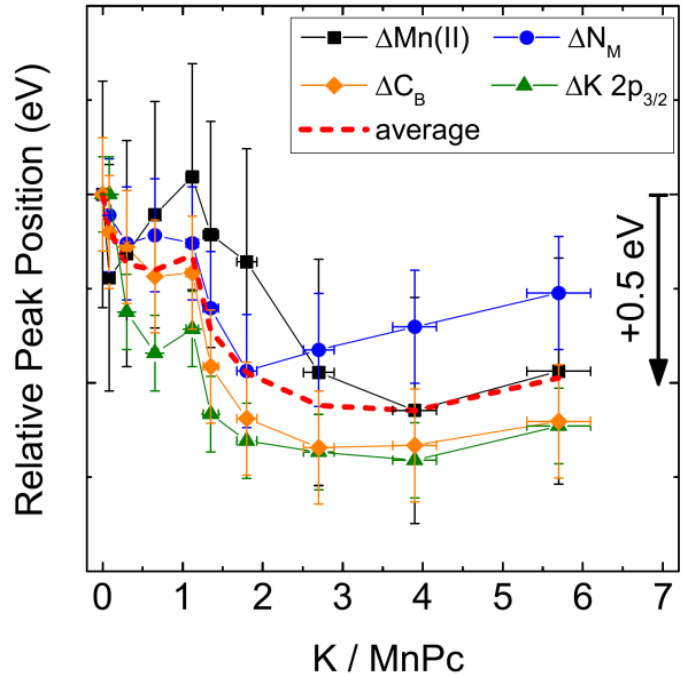

**Figure 4:** Core level peaks shifts as a function of K dosage. The relative positions for Mn(II) $2p_{3/2}$, $N_M$ 1s, $C_B$ 1s, and K $2p_{3/2}$ are according to Fig. 3. The red dashed line represents the average of these peak positions.

By comparing the spectra at the two excitation lines (150 eV and 21.2 eV) the first observation can be made for the pristine MnPc film. The H peak recorded at 150 eV excitation energy is more pronounced than the one at 21.2 eV excitation energy. This is due to the hybrid 3d character of the H feature which has a larger cross section at 150 eV excitation than the ligand related H−1 feature.[79]

The second observation is the additional feature assigned as A in Fig. 5 which arises at small amounts of K ($N_K$ = 0.3) and fades away above 1.8 K atoms per MnPc molecule. As feature A is not present in the 21.2 eV excitation energy spectra it could be assigned to a further Mn 3d related state at higher binding energy (∼ 2.5 eV). This assumption is strongly sustained by theoretical calculations by Brumboiu et al.,[48] who showed that different ground state electron occupation of the Mn 3d states provide different symmetry of the HOMO and different bond lengths in the MnPc molecular structure. According to them the most stable ground state configuration in thin MnPc films is $b_{1g}^0 a_{1g}^1 e_g^3 b_{2g}^1$ which does not have the additional A feature. However, the calculated electron configuration of $b_{1g}^0 a_{1g}^1 e_g^2 b_{2g}^2$ has a feature exactly at the energy position of peak A from Fig. 5. In conclusion, upon K intercalation the electronic configuration, molecular structure, and symmetry of the MnPc molecule is modified. As it will be shown later by NEXAFS measurements for $N_K < 1.8$, the donated electrons do not occupy empty Mn 3d states, which implies that initially the ligand related states are populated. The fade out of feature A for $N_K \sim 1.8$ is related to changes in the orbital configuration due to the occupation of other Mn 3d hybrid states by additional donated electrons.[58]

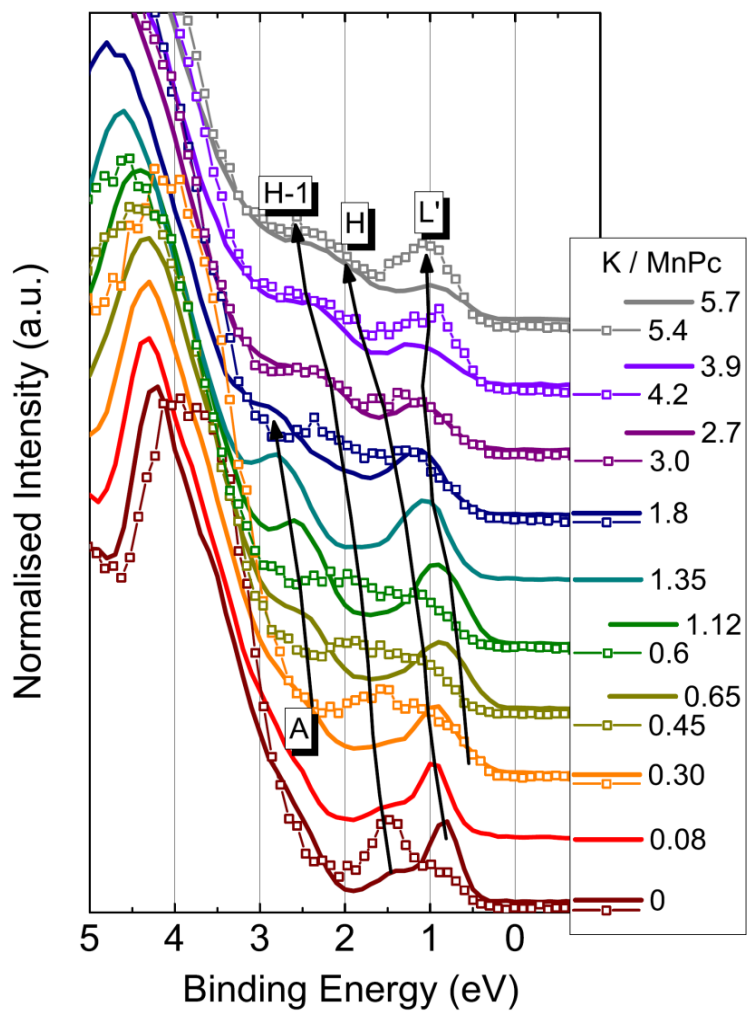

**Figure 5:** Evolution of the VB spectra as a function of the K amount. The solid lines represent spectra acquired at the photon excitation energy of 150 eV while the spectra with symbols are UPS measurements from Fig. 1(c) where the He I excitation line (21.2 eV) was employed. The features H−1, H, and L' represent the HOMO−1, HOMO, and filled LUMO states. Feature A is present only in the high excitation energy VB spectra. The vertical arrows mark the evolution of the peak positions.

The shifts of the VB features acquired at 150 eV excitation energy reproduce well the ones presented in the UPS−IPS measurements from Fig. 1. Moreover, the VB features recorded at 150 eV excitation energy are smeared out at high K dosages like in the UPS data.

## Near Edge X-ray Absorption Fine Structure

NEXAFS measurements at the manganese $L_{2,3}$, nitrogen $K$, and carbon $K$ edges were performed ,on pristine and K intercalated MnPc films. The spectra were recorded at normal ($\theta$ = 90°) and grazing ($\theta$ = 20°) angles of incidence with respect to the sample surface. Since the X-ray absorption phenomena indirectly provides information on the local unoccupied electronic states[80], the NEXAFS data were interpreted supplementary to the IPS results.

Due to the polycrystalline nature of the Co substrate and the fairly thick MnPc layer (10 nm) the molecules do not show highly preferential orientations. Only minor differences are present in the line shape and intensity of the spectra acquired at 90° (solid lines) or 20° (dashed lines) angles of incidence (Fig. 6). These suggest that, on average, the molecules are rather standing than lying.

Figure 6(a) illustrates the evolution of Mn $L_2$ (higher photon energy) and $L_3$ (lower photon energy) edges as a function of the intercalated K atoms. The two edges stem from electronic transitions from Mn $2p_{3/2}$ ($L_3$) and $2p_{1/2}$ ($L_2$) core levels onto unoccupied Mn 3d states, or in case of MnPc, onto hybridized 3d states.[81]

For pristine MnPc three features of the $L_3$ edge are visible, denoted as A, B, and C [Fig. 6(a)]. Up to 1.35 K atoms per molecule no significant changes are noticeable except of some slight shifts which can be correlated with the Mn 2p core level shifts from Fig. 3(a). However, above this threshold ($N_K \geq 1.8$) the $L_3$ absorption features undergo significant changes. Since the evolution of the features A, B, and C cannot be followed through the changes upon K doping the newly appeared features are denoted as D, E, and F. These undergo shifts of 0.5 eV toward lower energies. The striking development, however, is the rise of these peaks, especially feature D. The $L_2$ edge, on the other hand, remains almost constant in intensity only the spectral weight for the fully doped MnPc ($N_K$ = 5.7) is set on the first two features. These are more imposing as for the pristine material while the third feature fully disappeared.

Theoretical interpretation of the L absorption edge is rather complicated due to initial $3d^N$ and final $2p^53d^{N+1}$ states to which all the partially empty levels in the atom contribute. Moreover, the hybridization of Mn 3d states with N 2p states adds up to the difficulty. Theoretical calculations are available in literature, but often with contradictory results (see e.g. literature in Refs. 75, 81). Additionally, multi-electron effects further complicate the X-ray absorption spectra analysis.[47,80]

A comparison of the Mn $L_{2,3}$ edges lineshape from MnPc with XAS data on different inorganic Mn compounds[82,83] does not provide helpful information: the $L$ edge of the fully doped $K_{5.7}$[MnPc] film is similar to the one of $Mn^{2+}$ compounds, e.g. MnO (Ref. 83) or $MnF_2$ (Ref. 82), while the $L$ edge of pristine MnPc can be assessed with a superposition of XAS spectra of inorganic Mn compounds with $Mn^{3+}$ and $Mn^{4+}$ ionic behavior.[84,85] However, due to different final state effects, e.g. multiplet splitting,[47] the XAS line shapes are strongly influenced. Moreover, the influence of crystal field has to be taken into consideration, as with the change in its strength from 0 eV to 1.8 eV large modifications in the XAS spectra can be produced.[86] Thus, K intercalation strongly reduces the strength of the crystal field at the Mn site while it does not change the metal ion valence state.[17] Stepanow et al. concluded that the high-spin configuration (S = 5/2, with the orbital occupancy of $b_{1g}^0 a_{1g}^1 e_g^2 b_{2g}^1$ is the sole result of reduced crystal field strength.[17]

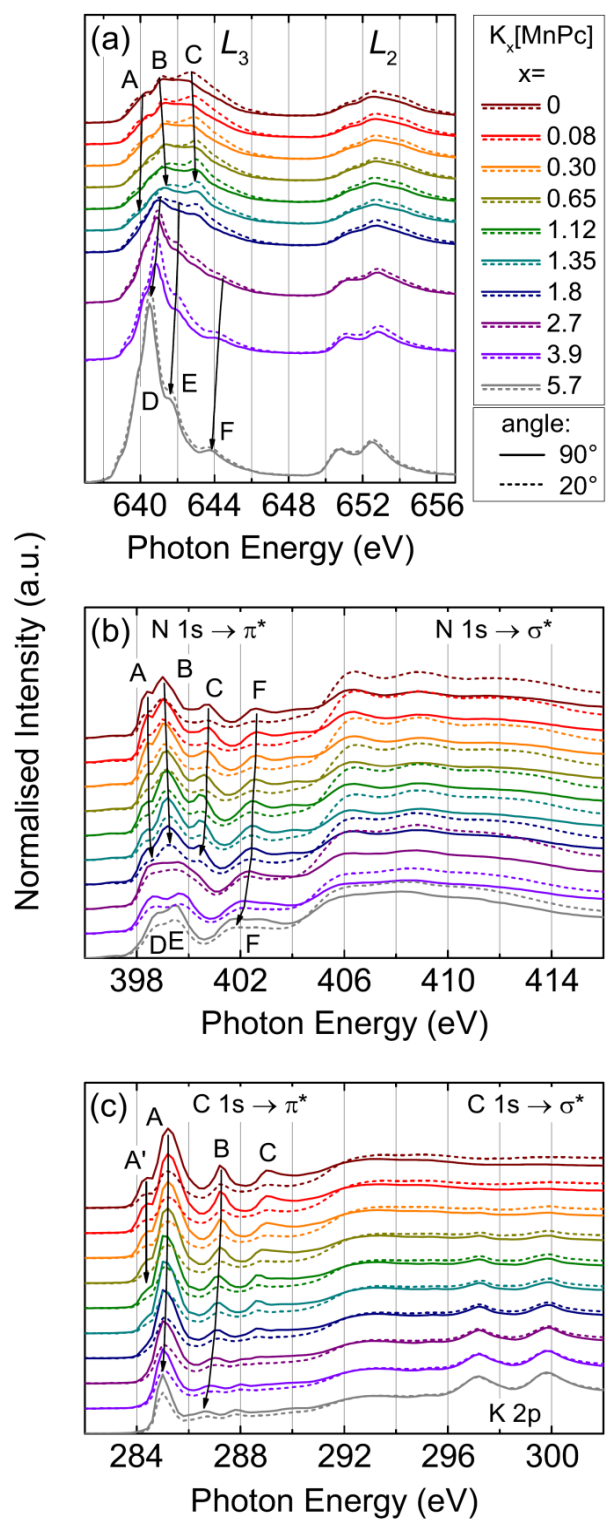

**Figure 6:** Evolution of the NEXAFS spectra (from top to bottom) with increasing amount of intercalated K per MnPc molecule. Measurements performed at normal (90°, solid lines) and grazing (20°, dashed lines) incidence of light with regard to the sample surface. The legend holds for all panels. (a) Spectra of Mn $L_{2,3}$ edges. The arrows track the peak positions evolution in the $L_3$ edge region (A, B, C → D, E, F). (b) N $K$ edge spectra: N 1s → $\pi^*$ and N 1s → $\sigma^*$ transitions regions. The arrows follow the evolution of the features labeled A, B, C, and F (A, B, C → D, E). (c) C K edge spectra: C 1s → $\pi^*$ and C 1s → $\sigma^*$ transitions regions. The arrows follow the evolution of the A', A, and B features.

Important information can be gained from the branching ratio (BR) of the $L_2$ and $L_3$ core-valence transitions.[87–89] The *BR* is defined as:

$$BR = \frac{A(L_3)}{A(L_2)+A(L_3)}, \qquad (3)$$

where $A(L_x)$ represents the area of the respective $L$ edge.[89] Thus, the *BR* for pristine MnPc and fully doped K$_{5.7}$[MnPc] was determined to be 0.68 ± 0.03 and 0.79 ± 0.03, respectively. The value for the pristine phase is just between the calculated[87,89] low- and high-spin states and attributed to the intermediate-spin state of S = 3/2 (Refs. 17, 27). From the increase of the *BR* upon K intercalation an increase of the spin state can be deduced,[89] therefore, K$_{5.7}$[MnPc] would have a high-spin state of S = 5/2 (Refs. 17, 23).

It has to be emphasized that the high-spin state of S = 5/2 can only be achieved if none of the donated electrons would occupy hybrid Mn 3d states and the original electron configuration would rearrange to $b_{1g}^0 a_{1g}^1 e_g^2 b_{2g}^1$ (Ref. 17). However, as previously reported by Mahns et al.[33] and observed in this work, donated electrons occupy and highly influence hybrid Mn 3d states. In consequence, the resulting spin state could only be as high as S = 2.

In the second set of NEXAFS measurements the evolution of the N K edge was recorded during K doping of the MnPc film. In Fig. 6(b) the spectra of N 1s → π* and N 1s → σ* transitions are plotted for both normal and grazing incidence of light. Starting with the pristine MnPc the evolution of four features (A, B, C, and F) can be followed. Up to $N_K$ = 1.8 the peaks A and B are unchanged, while C and F slightly shift toward lower energy. For $N_K$ > 1.8 feature F can be further followed with a total shift of 0.5 eV, however, A, B, and C vanish and for $N_K$ ≥ 3.9 two new features (D and E) arise.

K intercalated metal free phthalocyanine (H2Pc) illustrates only three features for the N 1s → π* transitions.[34] These comprise 1/4$^{th}$ of the LUMO, the rest being attributed to C 2p antibonding states. Moreover, the N *K* edge of CuPc is also composed of only three features[40,90] which suggests that the main contribution is given by the same π* empty states located around the pyrrole rings and composed of both N 2p and C 2p antibonding states. The N K and C K edges can be overlapped with relatively good agreement.[90] K addition to CuPc shows no spectacular changes, just the usual downward shift of the features.[40] However, FePc,[43] CoPc,[81] and MnPc,[75,81] present four distinct features, similar to the results of this work. It is argued that the first two features (A and B) belong to transitions from the two distinct nitrogen core levels (N$_M$ and N$_P$) into the same low lying empty $\pi^*$ state.[43] A second argument is that features A and B have significant contribution from the central metal ion, thus, a hybridization between N 2p and Mn 3d states occurs.[75,81] This argument is well supported by both the N $K$ edge evolution as well as by the Mn $L$ edges evolution. Thus, the empty electronic states are composed, to some extent, of hybrid 3d orbitals[78] highly influenced by electron donation from the intercalated K atoms. Furthermore, the findings by Mahns et al.[33] regarding the stable electronic configurations for 1, 2, and 4 K atoms per MnPc are well reproduced within the N $K$ edge evolution [Fig. 6(b)]. The features A, B, and C survive up to the net doping of $N_K$ = 1.8 (almost K2[MnPc]) and the rising features D and E are clearly distinguishable above $N_K$ = 3.9 (i.e. K$_4$[MnPc]). The spectrum at $N_K$ = 2.7 can be reproduced as a linear combination of the spectra with $N_K$ = 1.8 and $N_K$ = 3.9. Thus, the existence of the doped phases K$_2$[MnPc] and K$_4$[MnPc] were deduced and the instability of the one with 3 K atoms per MnPc was proven. The existence of the 3 stable electron doped configurations was determined by in situ spectroscopic ellipsometry measurements as well.[91]

A third set of NEXAFS spectra presents the evolution of the C K edge of MnPc via K doping [Fig. 6(c)]. Both C 1s → $\pi^*$ and C 1s → $\sigma^*$ transitions regions are marked in the figure. In the C 1s → $\sigma^*$ region also the evolution of the K 2p contribution can be found. The C 1s → $\pi^*$ region provides four distinct features: A', A, B, and C. Contrary to the N $K$ edge these features are present in H$_2$Pc,[34] and CuPc,[40,90] as well as in FePc,[43] or CoPc C $K$ edges.[42] Full agreement on the assignments of the first two peaks (A' and A) is provided in literature: A' and A are caused by excitation from benzene (C$_B$) and pyrrole (C$_P$) carbon atoms [see e.g. Fig. 3(d) and (e)], respectively, into the LUMO state.[34,40,42,43,90] The energy difference between A' and A features matches exactly the one between C$_B$ and C$_P$ features. Upon K addition peak A slightly shifts toward lower energies and decreases in intensity while feature A' only decreases in intensity and fully disappears for $N_K$ > 1.8. The disappearance of A' is attributed to the filling of the LUMO with donated K 4s electrons. Both features A and B have a shift of 0.5 eV toward lower energies. All shifts of the NEXAFS features toward lower energies can be rationalized as the outcome of the electronic structures relaxation by filling up the unoccupied states with donated electrons.

In conclusion, the spectral evolution of Mn $L_{2,3}$, N $K$, and C $K$ edges upon K doping of MnPc provides information where the donated e$^-$ most probably reside on the molecule. For $N_K$ = 1 the e$^-$ cloud covers the whole molecule [Fig. 3(f)]. The second donated e$^-$ (for $N_K$ = 2) is located in the center of the MnPc where it occupies hybridized Mn 3d and N 2p empty states [Fig. 3(g)]. These states are further influenced and spread towards N 2p up to $N_K$ = 4 which is schematically represented in Fig. 3(h).

## Conclusions

To summarize, the impact of K addition on the occupied and unoccupied electronic states of MnPc films was studied by combined UPS and IPS measurements. Moreover, the core level spectra were determined by XPS while NEXAFS spectroscopy was employed for a detailed study of the unoccupied states.

The UPS-IPS results present a shift towards higher BE of the bands upon K intercalation. Though a rigid shift is expected according to literature,[57] slight differences between the total shift of the occupied (~ 0.8 eV) and unoccupied (~ 1.2 eV) electronic states were observed. Moreover, the shift of the core levels is even less (~ 0.5 eV), implying a more complicated scenario than a simple Fermi level shift. The influence of the photo-induced core hole on the valence states is larger than on the stronger bound deeper levels.

From UPS-IPS measurements not only the previously determined transport bandgap of (1.2 ± 0.3) eV (Ref. 49) was deduced but also its evolution upon n-type doping was studied. It was presented before that $O_2$ exposure (p-type doping) of the MnPc film influences the HOMO state and increases the transport bandgap to (1.45 ± 0.30) eV (Ref. 49). Upon K intercalation, however, the LUMO state is filled up with electrons and leads to the appearance of the L' feature in the HOMO region. As all the electronic levels shift and the LUMO state gradually disappears upon K intercalation, it is difficult to give a precise value of the transport bandgap. Its value, however, does not decrease below (0.9 ± 0.3) eV due to the Fermi level pinning.

XPS provides information on the possible positioning of the $K^+$ ions while NEXAFS offers evidence where their donated electrons reside. The ions have a large impact on the spectral shape of the evolution of the C and N core levels upon doping, while the Mn $2p_{3/2}$ spectrum is not affected. Thus the $K^+$ ions prefer the positions near the asa-bridging nitrogen (i.e. $N_P$) and on the benzene rings. According to literature the crystalline structure of the pristine MnPc film provides a stacking of the molecules with consecutive pyrrole N atoms above the Mn central atoms.[23,28] In the AM intercalated film it was reported that $K^+$ ions prefer at first the location near $N_P$ (Refs. 23,43) and only after the saturation ($N_K$ = 4), the benzene carbon atoms were influenced.

The C K edge was constantly shifted upon K intercalation. The Mn $L$ edge evolution on the other hand has a sudden jump between $K_1$[MnPc] and $K_2$[MnPc] which implies that the second donated electron per MnPc molecule already occupies the hybridized antibonding states of the Mn center. Moreover, the N $K$ edge spectrum changes dramatically between $K_2$[MnPc] and $K_4$[MnPc] excluding the possibility of a state in between. This also denotes the implication of N antibonding states in accepting the third and fourth donated electrons. Moreover, the sudden spectral changes observed in the evolution of the Mn $L$ and the N $K$ edges evidences the strong hybridization of the Mn 3d and the N 2p states.[75] Though not large, the angular dependence of N $K$ and C $K$ edges upon K intercalation denote a modification in symmetry of the unoccupied electronic states. The branching ratio of the Mn $L_{2,3}$ edges indicates a modification of the total spin of the MnPc molecule upon doping from S = 3/2 to a higher one (S = 2 or even S = 5/2).

## Acknowledgement

This work was supported by the Deutsche Forschungsgemeinschaft (DFG) Research Unit FOR 1154 "Towards Molecular Spintronics". The authors acknowledge Helmholtz-Zentrum Berlin (HZB) for the allocation of synchrotron radiation beamtime and financial support.